\documentclass[aoas,MSNbibl,nameyear,seceqn,dvips]{arximspdf}
\usepackage{mathbh}
\usepackage{graphicx}

\doi{10.1214/12-AOAS569} 
\volume{6}
\issue{4}
\pubyear{2012}
\firstpage{1795}
\lastpage{1813}



\setattribute{abstract}{width}{300pt}

\begin{document}
\begin{frontmatter}

\title{Latent demographic profile estimation in hard-to-reach groups}
\runtitle{Latent demographic profile estimation}

\begin{aug}
\author[A]{\fnms{Tyler H.} \snm{McCormick}\corref{}\thanksref{t1}\ead[label=e1]{tylermc@u.washington.edu}}
\and
\author[B]{\fnms{Tian} \snm{Zheng}\thanksref{t2}\ead[label=e3]{tzheng@stat.columbia.edu}\ead[label=u1,url]{http://www.stat.columbia.edu/\textasciitilde tzheng}}
\runauthor{T. H. McCormick and T. Zheng}
\affiliation{University of Washington and Columbia University}
\address[A]{Department of Statistics\\
Department of Sociology\\
Center for Statistics and the Social Sciences\\
University of Washington\\
Box 354320 Seattle, Washington 98195\\
USA\\
\printead{e1}}
\address[B]{Department of Statistics\\
Columbia University\\
1255 Amsterdam Ave.\\
New York, New York 10027\\
USA\\
\printead{e3}\\
\printead{u1}} 
\end{aug}

\thankstext{t1}{Supported in part by NIAID Grant R01 HD54511. This work was partially
completed while Tyler McCormick was supported by a Google Ph.D.
fellowship in Statistics.}

\thankstext{t2}{Supported by NSF Grant SES-1023176 and a Google
faculty research award.}

\received{\smonth{2} \syear{2011}}
\revised{\smonth{2} \syear{2012}}

%
\begin{abstract}
The sampling frame in most social science surveys excludes members of
certain groups, known as \textit{hard-to-reach groups}. These groups, or
subpopulations, may be difficult to access (the homeless, e.g.),
camouflaged by stigma (individuals with HIV/AIDS), or both (commercial
sex workers). Even basic demographic information about these groups is
typically unknown, especially in many developing nations. We present
statistical models which leverage social network structure to estimate
demographic characteristics of these subpopulations using Aggregated
relational data (ARD), or questions of the form ``How many X's do you
know?'' Unlike other network-based techniques for reaching these
groups, ARD require no special sampling strategy and are easily
incorporated into standard surveys. ARD also do not require respondents
to reveal their own group membership. We propose a Bayesian
hierarchical model for estimating the demographic characteristics of
hard-to-reach groups, or \textit{latent demographic profiles}, using ARD.
We propose two estimation techniques. First, we propose a Markov-chain
Monte Carlo algorithm for existing data or cases where the full
posterior distribution is of interest. For cases when new data can be
collected, we propose guidelines and, based on these guidelines,
propose a simple estimate motivated by a missing data approach. Using
data from McCarty et al. [\textit{Human Organization} \textbf{60}
(2001) 28--39], we estimate the age and gender profiles of six
hard-to-reach groups, such as individuals who have HIV, women who were
raped, and homeless persons. We also evaluate our simple estimates
using simulation studies.
\end{abstract}

%
\begin{keyword}
\kwd{Aggregated relational data}
\kwd{hard-to-reach populations}
\kwd{hierarchical model}
\kwd{social network}
\kwd{survey design}
\end{keyword}

\end{frontmatter}

\section{Introduction}\label{secintro}
Standard surveys often exclude members of the certain groups, know as
\textit{hard-to-reach groups}. One reason these individuals are excluded
is difficulty\vadjust{\goodbreak} accessing group members. Persons who are homeless are
very unlikely to be reached by a survey which uses random-digit
dialing, for example. Other individuals can be accessed using standard
survey techniques, but are excluded because of issues in reporting.
Members of these groups are often reluctant to self-identify because of
social pressure or stigma [\citet{Shelley1995}]. Individuals who
are homosexual, for example, may not be comfortable revealing their
sexual preferences to an unfamiliar survey enumerator. A third group of
individuals is difficult to reach because of issues with both access
and reporting (commercial sex workers, e.g.).

Even basic demographic information about these groups is typically
unknown, especially in developing nations. We propose a Bayesian
hierarchical model for estimating the demographic characteristics of
hard-to-reach groups, or \textit{latent demographic profiles}.
Specifically, these profiles reveal features such as the number of
males in a certain age range, say, 20--30 years old, who have HIV.
Sociologically, this information yields insights into the
characteristics of some of the most socially isolated members for the
population. Along with its contribution to our understanding of
contemporary social institutions, estimating demographic profiles for
these groups also has public health benefits. The distribution of
infected individuals influences the size of the public health response.
UNAIDS---the joint United Nations program on HIV/AIDS, for example,
currently sponsors several projects using a variety of techniques to
estimate the sizes of populations most at-risk for
HIV/AIDS [\citet
{UNAIDS03}]. The proposed method would, along with estimating the size
of the population, provide latent demographic profiles. This
information would not only help calibrate the scale of the response but
also tailor programs to the specific needs of population members.

One approach to estimating demographic information about hard-to-reach
groups is to reach members of these groups through their social
network. Some network-based approaches, such as Respondent-driven
Sampling (RDS), recruit respondents directly from other respondents'
networks [Heckathorn (\citeyear{DH97,DH02})], making the sampling mechanism
similar to a stochastic process on the social network [\citet
{Goel2009}]. RDS affords researchers face-to-face contact with members
of hard-to-reach groups, facilitating exhaustive interviews and even
genetic or medical testing. The price for an entr\'ee to these groups
is high, however, as RDS uses a specially designed link-tracing
framework for sampling. Estimates from RDS are also biased because of
the network structure captured during selection, with much statistical
work surrounding RDS being intended to reweight observations from RDS
to have properties resembling a simple random sample.

Another approach is Aggregated relational data (ARD) or ``How many X's
do you know'' questions [\citet{PKetal98eval}]. In these questions,
``X'' defines a population of interest (e.g., How many people who are
homeless do you know?). A~specific definition of ``know'' defines the
network the respondent references when answering the question. In
contrast to RDS, ARD do not require reaching members of the
hard-to-reach groups directly. Instead, ARD access hard-to-reach
groups indirectly through the social networks of respondents on
standard surveys. ARD never affords direct access to members of
hard-to-reach populations, making the level of detail achievable though
RDS impossible with ARD. Unlike RDS, however, ARD require no special
sampling techniques and are easily incorporated into standard surveys.
ARD are, therefore, feasible for a broader range of researchers across
the social sciences, public health, and epidemiology to implement with
significantly lower cost than RDS.

In this paper, we propose a model for estimating latent demographic
profiles using ARD. The ease of implementation of ARD means that the
models proposed here will make the demographic characteristics of
hard-to-reach groups available to the multitude of researchers
collecting data using standard survey methodology. Specifically, we
propose a Bayesian hierarchical model for estimating the demographic
characteristics of hard-to-reach groups using ARD. When the full
posterior is of interest, we propose a Markov-chain Monte Carlo algorithm.

Given the ease of collecting ARD, we speculate that many researchers
may be interested in including ARD questions on future surveys. In this
case, we show that estimates for some network features very close to
those achieved using MCMC can be obtained using significantly simpler
estimation techniques under certain survey design conditions. Along
with giving survey guidelines, we propose a simpler estimation
technique based on the EM algorithm and regression. Using data
from \citet{CMetal01}, we estimate the age and gender profiles of
six hard-to-reach groups, such as individuals who have HIV, women who
were raped, and homeless persons.

%

In Section~\ref{secprev} we contextualize our proposed method by
reviewing previous statistical methods for estimating network features
using ARD. Then, we describe a method for estimating demographic
profiles from hard-to-reach populations. Section~\ref{secdata}
illustrates our method using data from \citet{CMetal01}. After
demonstrating the utility of our model, Section~\ref{secsimple}
describes how, under certain survey design conditions, we can obtain
similar estimates without the computational sophistication required by MCMC.

\section{Previous research on ARD}
\label{secprev}
ARD are commonly used to estimate the size of populations that are
difficult to count directly. The scale-up method, an early method
for ARD, uses ARD questions where the subpopulation size is known
(people named Michael, e.g.) to estimate degree
in a straightforward manner. Suppose that you know two persons
named Nicole, and that at the time of the survey, there were 358,000
Nicoles out of 280 million Americans. Thus, your two Nicoles
represent a fraction ($2/358$,000) of all the Nicoles. Extrapolating
to the entire country yields an estimate of
($2/358$,000)${}\times{}$(280 million)${}={}$1560 people known by you. Then,
the size of unknown subpopulations is estimated by solving the given
equation for the unknown subpopulation size with the estimated
degree. Using this method, ARD has been used extensively to
estimate the size of populations such as those with HIV/AIDS,
injection drug users, or the homeless [e.g., Killworth et~al.
(\citeyear{PKetal90,PKetal98soc})].



The scale-up method is easy to implement but does not account for network
structure. Consider, for example, asking a respondent how many
people named ``Rose'' she/he knows. If knowing someone named Rose
were entirely random, then each respondent would be equally likely
to know each of the one-half million Rose's on the hypothetical
list; that is, each respondent on each Rose is a Bernoulli trial
with a fixed success probability. Network structure makes these types
of independence assumptions
invalid. Since Rose is most common among older
females and people are more likely to know individuals of similar
age and the same gender, older female respondents are more likely to
know a
given Rose than older male respondents. Assuming independent responses
induces bias in the individuals'
responses. Since estimates of hard-to-count populations are then
constructed using responses to ARD,
the resulting estimates are also biased [\citet{PKetal98eval,BJKR1991}].

%
\citet{TZetal06} and \citet{MSZ08} propose hierarchical
models for ARD which partially address the manifestations of network
structure present in ARD. \citet{MSZ08} develop a model
specifically for estimating respondents' degree (network size) and
population degree distribution. Though this model accounts for the
network structure described in the above example, \citet{MSZ08} do
not address hard-to-reach groups. \citet{TZetal06} present a
model which estimates the sizes of hard-to-reach groups [see Figure 5
in \citet{TZetal06}]. This paper presents a model which provides
richer information about hard-to-reach groups by estimating both
subpopulation sizes and the demographic breakdown of individuals within
these groups.

\section{Estimating latent profiles}\label{secmodel}

In this section we describe a model for estimating latent demographic
profiles for hard-to-reach groups. This method will provide information
about the demographic makeup of groups which are often difficult to
access using standard surveys, such as the proportion of young males
who are infected with HIV. The observations, $y_{ik}$, represent the
number of individuals in subpopulation $k$ known by respondent $i$. In
ARD, respondents are conceptualized as \textit{egos}, or senders of ties
in the network. We divide the egos into groups based on their
demographic characteristics (males 20--40 years old, e.g.). The
individuals who comprise the counts for ARD are the \textit{alters}, or
recipients of links in the network. The alters are also divided into
groups, though the groups need not be the same for both the ego and the
alter groups. Under this setup members of hard-to-reach groups are one
type of alter. Thus, determining the alter groups determines the
demographic characteristics of the hard-to-reach groups which can be estimated.
%
%
%
%
We model the number of people that respondent $i$ is connected to in
subpopulation $k$ as
%
\begin{eqnarray}\label{eqrecallmodel}
y_{ik} &\sim& \mbox{Neg-Binom}
(\mu_{ike}, \omega_k)
\nonumber\\[-8pt]\\[-8pt]
&&\eqntext{\mbox{where }\displaystyle
\mu_{ike} = d_i \sum_{a=1}^A
m(e, a) h(a, k)}
\end{eqnarray}
and $\omega_k$ represents the
variation in the relative propensity of respondents within an ego
group to form ties with individuals in a particular subpopulation
$k$.
The degree of person $i$ is $d_i$ and $e$ is the ego group that
person $i$ belongs to. The $h(a,k)$ term is the relative size of
subpopulation group
$k$ within alter group $a$ (e.g., 4\% of males between ages 21 and 40
are named Michael). The mixing coefficient, $m(e,a)$, for a respondent
with degree $d_i=\sum_{a=1}^A d_{ia}$ between ego-group $e$ and
alter-group $a$ is
\[
m(e,a) = \mathrm{E}
\biggl(\frac{d_{ia}}{d_i}\Big\vert i\mbox{ in ego group }e
\biggr),
\]
where $d_{ia}$ is the number of person $i$'s acquaintances in alter
group $a$. That is, $m(e,a)$ represents the expected fraction of the
ties of someone in ego-group $e$ that go to people in alter-group
$a$. For any group $e$, $\sum_{a=1}^{A} m(e,a) = 1$.

The vector of mixing rates for an ego group, $(m(e,1),\ldots,
m(e,A))^{T}$, enters the likelihood via an inner product with $h(a,k)$;
therefore, its components are only identifiable if the $A$ by $K$
matrix of $h(a,k)$ terms, $\mathbf{H}_{A\times K}$, has rank $A$. This
condition requires $K > A$ and that the columns of $\mathbf{H}_{A
\times K}$ not be perfectly correlated. When all elements of $\mathbf
{H}_{A \times K}$ are fixed, then (\ref{eqrecallmodel}) is the LRNM
model from \citet{MSZ08}. Specifically, \citet{MSZ08} propose
asking ARD questions about populations where the elements of $h(a,k)$
are readily available, such as first names in the United States
population. When $h(a,k)$ is known, it is simply $N_{ak}/N_{a}$ or the
number of individuals in alter group $a$ who have characteristic $k$
divided by the number of individuals in alter group $a$.

In hard-to-reach groups, $h(a,k)$ is rarely known. In many cases, even
the number of individuals in a hard-to-reach group, $N_{k}$, is
unknown. In the following section we propose a method for estimating
$h(a,k)$ for hard-to-reach groups using information from groups when
$h(a,k)$ is available. This method provides information beyond the size
of the subpopulation group, also estimating the number of individuals
in each of the $a$ alter groups, $N_{ak}$.

In summary, the number of people that person $i$ knows in subpopulation~$k$,
given that person $i$ is in ego-group $e$, is based on person $i$'s
degree ($d_i$), the proportion of people\vadjust{\goodbreak} in alter-group $a$ that belong
to subpopulation $k$, ($h(a,k)$), and the mixing rate between people
in group $e$ and people in group $a$, ($m(e,a)$). Additionally, if
we observe random mixing, then $m(e,a)=N_{a}/N$.

Similar to \citet{TZetal06}, a negative binomial model is assumed
in (\ref{eqrecallmodel}) for each $y_{ik}$ with an overdispersion,
$\omega_{k}$, parameter measuring the residual relative propensity of
respondents to form ties with individuals in group~$k$, controlling for
the variations that define the ego groups.

%

\subsection{Latent demographic profiles}
\label{secnewmodel}

We propose a two-stage estimation procedure. We first use a multilevel
model and Bayesian inference to estimate~$d_{i}$, $m(e,a)$, and $\omega_{k}'$ using the
latent nonrandom mixing model described in McCormick, Salganik and Zheng
(\citeyear{MSZ08}) for the
subpopulations where $h(a,k)=N_{ak}/N_{a}$ is known. Second,
conditional on this information, we estimate the latent profiles for
the remaining subpopulations.

For the estimation of the \citet{MSZ08} model components, we
assume that $\log(d_{i})$
follows a normal distribution with mean $\mu_{d}$ and standard
deviation $\sigma_{d}$. \citet{TZetal06} postulate that this prior
should be reasonable based on previous work, specifically \citet
{CMetal01}, and found that the sampler described using this prior mixed
well and satisfied posterior predictive checks. \citet{MSZ08}
also conducted simulation experiments which demonstrated that the shape
of the posterior was not an artifact of this prior assumption. We
estimate a value of $m(e,a)$ for all $E$ ego groups and all $A$
alter groups. For ego group, $e$, and alter
group, $a$, we assume that $m(e,a)$ has a normal prior distribution
with mean
$\mu_{m(e,a)}$ and standard deviation $\sigma_{m(e,a)}$.
For $\omega_{k}'$, we use independent $\operatorname{uniform}(0,1)$ priors on the inverse
scale, $p(1/\omega_{k}')\propto1$. Since $\omega_{k}'$ is constrained
to $(1,\infty)$, the inverse falls on $(0,1)$. The Jacobian for the
transformation is $\omega_{k}'^{-2}$.
For the latent profiles, define $\mathbh 1_{h(a,k)}$ as
the indicator of the latent profiles. The matrix $h(a,k)$ is defined as
$N_{ak}/N_a$ when population information is available ($\mathbh1_{h(a,k)}=0$) and entries to be estimated
($\mathbh1_{h(a,k)}=1$) are given normal priors on the log scale with
mean $\mu_{h}$ and standard deviation $\sigma_{h}$. That is, we model
each $\log(h(a,k))\sim\mbox{N}(\mu_{h}, \sigma^{2}_{h})$ with a common
mean and variance for all entries in the latent profile matrix. Since
many of the profiles are close to zero, we found that the additional
structure from a common prior across all entries improved convergence
without being too rigid to capture fluctuations in latent intensity.
Finally, we give noninformative uniform priors to the hyperparameters
$\mu_{d}$, $\mu_{m(e,a)}$, $\mu_{h}$, $\sigma_{d}$ and $\sigma_{m(e,a)}$, $\sigma_{h}$. The
joint posterior density can then be expressed as
\begin{eqnarray*}
&&
p\bigl(d,m(e,a),\omega',\mu_{d},
\mu_{m(e,a)},\sigma_{d},\sigma_{m(e,a)}|y\bigr)\\
&&\qquad\propto
\prod_{k=1}^{K} \prod
_{i=1}^{N}\pmatrix{y_{ik}+\xi_{ik}-1
\cr\xi_{ik}-1} \biggl(\frac{1}{\omega_{k}'} \biggr)^{\xi_{ik}}
\biggl(\frac{\omega_{k}'-1}{\omega_{k}'} \biggr)^{y_{ik}}
\\
&&\qquad\quad\hspace*{29.1pt}{} \times \biggl(\frac{1}{\omega_k} \biggr)^{2}\prod
^{N}_{i=1}N\bigl(\log(d_{i})|
\mu_{d},\sigma_{d}\bigr)
\\
&&\qquad\quad\hspace*{0pt}{}\times \prod_{e=1}^{E}N\bigl(m(e,a)|
\mu_{m(e,a)},\sigma_{m(e,a)}\bigr)
\\
&&\qquad\quad\hspace*{0pt}{}\times \mathbh1_{h(a,k)} \prod
_{k=1}^{K}\prod_{a=1}^{A}N
\bigl(h(a,k)|\mu_{h(a,k)},\sigma_{h(a,k)}\bigr),
\end{eqnarray*}
where $\xi_{ik}= d_i f (\sum_{a=1}^A   m(e, a)
h(a,k) )/(\omega_{k}'-1)$.

Adapting \citet{TZetal06} and \citet{MSZ08}, we use a
Gibbs--Metropolis algorithm in each iteration~$v$:
\begin{longlist}[(11)]
\item[(1)] For each $i$, update $d_{i}$ using a Metropolis step with
jumping distribution $\log(d^{*}_{i})\sim N(d_{i}^{(v-1)}\mbox{, (jumping
scale of $d_{i}$)}^{2})$.

\item[(2)] For each $e$, update the vector $m(e,\cdot)$ using a Metropolis
step. Define the proposed value using a random direction and jumping
rate. Each of the $A$ elements of $m(e,\cdot)$ has a marginal jumping
distribution
$m(e,a)^{*}\sim N(m(e,a)^{(v-1)}\mbox{, (jumping scale
of $m(e,\cdot)$)}^2)$.
Then, rescale so that the row sum is one.
%

\item[(3)] Update $\mu_{d}\sim N(\hat{\mu}_{d},\sigma^{2}_{d}/n)$, where
$\hat{\mu}_{d}=\frac{1}{n}\Sigma_{i=1}^{n}d_{i}$.

\item[(4)] Update $\sigma_{d}^{2}\sim
\mbox{Inv-}\chi^{2}(n-1,\hat{\sigma}_{d}^{2})$, where
$\hat{\sigma}_{d}^{2}= \frac{1}{n} \times
\Sigma_{i=1}^{n}(d_{i}-\mu_{d})^{2}$.

\item[(5)] Update $\mu_{m(e,a)}\sim N(\hat{\mu}_{m(e,a)},\sigma^{2}_{m(e,a)}/A)$, for each $e$ where
$\hat{\mu}_{m(e,a)}=\frac{1}{A}\Sigma_{a=1}^{A}m(e,a)$.

\item[(6)] Update $\sigma_{m(e,a)}^{2}\sim
\mbox{Inv-}\chi^{2}(A-1,\hat{\sigma}_{m(e,a)}^{2})$, for each $e$
where $\hat{\sigma}_{m(e,a)}^{2}= \frac{1}{A} \times
\Sigma_{a=1}^{A}(m(e,a)-\mu_{m(e,a)})^{2}$.

\item[(7)] For each $k$ with a known profile, update $\omega_{k}'$ using a
Metropolis step with
jumping distribution $\omega'^{*}_{k} \sim
N(\omega_{k}'^{(v-1)}\mbox{, (jumping scale of $\omega_{k}'$)}^{2})$.

We now proceed to estimate the $H$ latent profiles:%
\item[(8)] For each element of $h(a,k)$ where $\mathbh1_{h(a,k)}=1$, update
$h(a,k)$ using a Metropolis step with jumping distribution
$h(a,k)^{*}\sim N(h(a,k)^{(v-1)}$, (jumping scale of $h(a,k))^{2})$.

\item[(9)] Update $\mu_{h}\sim N(\hat{\mu}_{h},\sigma^{2}_{h}/(A\times H))$
for each $k$ where
\[
\hat{\mu}_{h}=\frac{1}{(A \times H)}\*\sum_{k=1}^{K}\sum_{a=1}^{A}
\mathbh1_{h(a,k)}h(a,k).
\]

\item[(10)] Update $\sigma_{h}^{2}\sim
\mbox{Inv-}\chi^{2}((A\times H)-1,\hat{\sigma}_{h}^{2})$
where
\[
\hat{\sigma}_{h}^{2}= \frac{1}{A\times H}
\sum_{k=1}^{K}\sum_{a=1}^{A}\mathbh1_{h(a,k)}\bigl(h(a,k)-\mu_{h}\bigr)^{2}.
\]

\item[(11)] For each $k$ where $h(a,k)$ is estimated, update $\omega_{k}'$
using a Metropolis step with
jumping distribution $\omega'^{*}_{k} \sim
N(\omega_{k}'^{(v-1)}$, (jumping scale of $\omega_{k}'$)$^{2}$).
\end{longlist}
Having $h(a,k)$ for some subpopulations is critical to estimating
latent structure through latent profiles. Often, $h(a,k)$ can be
obtained from publicly available sources (Census Bureau, Social
Security Administration, etc.) for subpopulations such as first names.
The number of populations with known $h(a,k)$ impacts the precision of
the estimates for subpopulations with unknown profiles. Adding another
known subpopulation increases the hypothetical sample size of each
question, in essence asking each respondent if they know more alters.
\citet{MSZ08} show that the total size of the subpopulations
asked is related to the variance of estimated degree. Since known
subpopulations are used to estimate degree, adding another
subpopulation impacts variability in degree estimation in the first
stage of our procedure, which propagates to estimates of $h(a,k)$. The
alter groups where information is available for known $h(a,k)$ also
limit the type of latent structure that can be estimated. \citet
{MSZ08} create alter groups based on age and gender but note that
separating alters based on other factors (such as race) would provide
valuable information. The Census Bureau collects the information
required to conduct such an analysis; however, \citet{MSZ08}
report that their efforts to obtain the data were ultimately unsuccessful.

The choice of populations with known $h(a,k)$ is also important in
ensuring that the mixing matrix is estimated appropriately. First, the
subpopulations with known sizes need to be sufficiently heterogeneous
with respect to their interactions with the ego groups to adequately
estimate the mixing matrix. If, for example, our mixing matrix consists
of only gender and we chose to use first names for subpopulations with
known $h(a,k)$, then we should use a set of both male and female names.
If we only asked male names, then we could estimate the propensity for
males/females to interact with males but could not estimate the
propensity of either gender to interact with females. Second, we make
an assumption about the representativeness of respondents' networks
rather than of the respondents themselves. For our method it would not
be an issue, for example, if we recruit a smaller fraction of men into
the survey than the proportion of men in the population. Instead, we
would encounter bias if the networks of the men we selected were not
representative of male networks in the population. This could happen,
for example, if we recruit only men who know a disproportionately large
number of women.
This issue could also be exacerbated by differential nonresponse.
Consider, for example, the case where individuals who know members of
the hard-to-reach groups are less likely to answer questions than the
general population. We continue this discussion in Section \ref
{secdata} where we postulate that errors in the estimates obtained in
our data could be from bias in the estimates of the mixing matrix.\vadjust{\goodbreak}

Finally, certain types of bias which are consistently associated with
ARD should also be considered when selecting the subpopulations with
known $h(a,k)$. We assume, for example, that the responses are free
from transmission error, when a respondent knows a member of a
subpopulation but is unaware of the alter's membership. \citet
{MSZ08} suggest using first names since they represent the minimum
conceivable possibility of transmission error. We also assume that
respondents accurately recall the number of individuals they know in a
given subpopulation. In reality, underestimation is common in large
groups [see \citet{McCormick2007wo} for a detailed discussion].

%
\section{Results for hard-to-count populations}
\label{secdata}
We use data from a telephone survey by \citet{CMetal01} with 1375
respondents and twelve names with known demographic profiles. These
data have been analyzed in several previous studies and are typical ARD
which are becoming increasingly common. The age and gender profiles of
the names are available from the Social Security Administration. On
this survey, ``know'' is defined ``that you know them and they know you
by sight or by name, that you could contact them, that they live
within the United States, and that there has been some contact (either
in person, by telephone, or mail) in the past 2 years.''
We then estimate latent profiles for seven subpopulations. Six are
groups often considered hard-to-count while the seventh uses ARD to
learn about population social structure.
%
\begin{figure}

\includegraphics{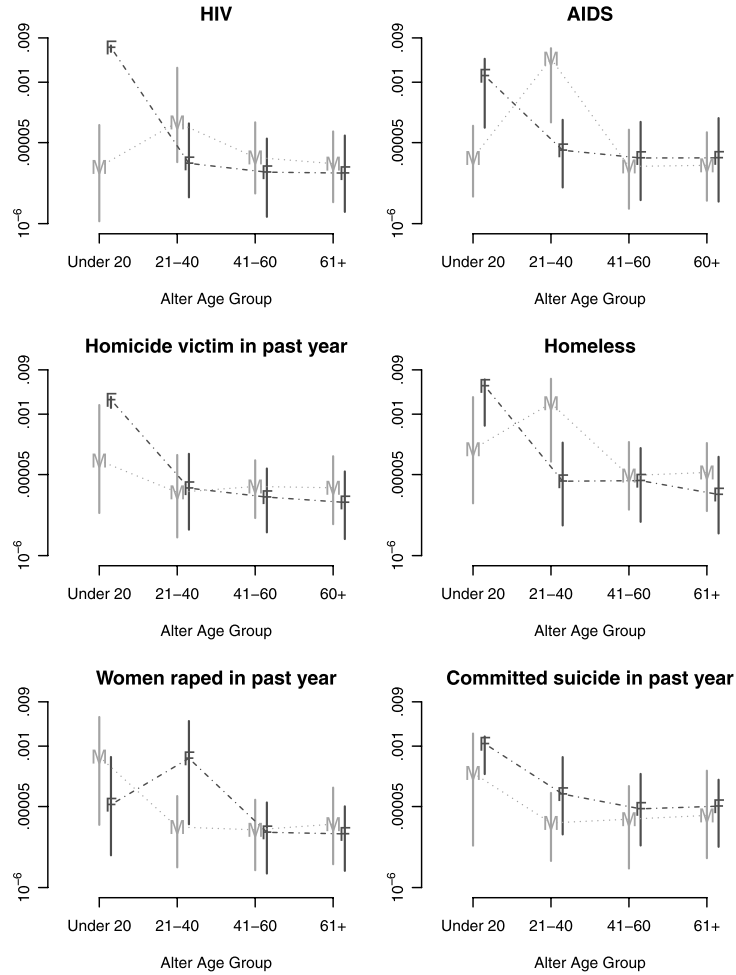}

\caption{Estimates of latent profiles for six hard-to-reach
populations. The lighter text represents males and the darker text
females. Letters correspond to posterior medians, while lines represent
the width of the middle half of the posterior distribution. The
estimated profiles are consistent with contemporary understanding of
the profiles of these groups.}
\label{fighidden1}
\end{figure}
Figure~\ref{fighidden1} displays the latent profiles for six
populations which are often described as hard-to-count. For both
individuals with HIV and those with AIDS we estimate the highest
concentration to be among youth and young adult respondents. We
estimate a higher concentration of young adult males than females for
both HIV and AIDS with the concentration decreasing with age.

Subpopulations such as victims of homicide or persons who have
committed suicide portray a key advantage of using ARD for measuring
these populations. Our model estimates characteristics of these
populations without requiring members of these populations to be
reached directly through our survey. We compared our estimates of the
number of individuals murdered in the past year with the 1999
\textit{Uniform Crime Reports} (\textit{UCR}) [\citet{UCR99}] and figures from the
Centers for Disease Control National Center for Injury Prevention and
Control (CDC) [\citet{CDC}]. A technical distinction between the
two sources for external validation is that the CDC figures measure
homicides (killing of another person) while UCR tally murders (unlawful
killing of another person). So-called justifiable homicides (police
officers using deadly force, e.g.) are therefore not counted in
the UCR figures. This distinction accounts for part of the discrepancy
between the two data sources (the FBI only keeps records on
firearms-related justifiable homicides), though the exact amount could
not be determined from available data. Also, the \citet{CMetal01}
survey took place partially in January of 1999 and partially in June,
meaning that this report does not capture precisely the period
respondents were asked to recall. Since homicide statistics do not
typically change drastically on a national scale over the course of a
year, we expect, nonetheless, that these figures are reasonable for
comparison. In all six age-gender categories, the UCR and CDC estimates
are within the middle 50$\%$ of the posterior distribution of our
estimates [computed by multiplying $h(a,k)$ by the number of
individuals in the given age-gender group]. For males 20--40, for
example, the UCR counts approximately 5300 murders while that CDC
counts just under 7300 homicides. Our method estimates the first
quartile of the posterior distribution as roughly 300 murders and the
third quartile as around 7300. Similarly, for females between 40 and
60 the middle half of our posterior lies between around 100 and 2300
while the UCR records around 700 and the CDC counts about 1900.
Overall our estimates underrepresent the disparity in the proportion of
male and female homicide victims, which we believe is due to the
individuals who are most likely associated with murdered individuals
being underrepresented in the survey frame. \citet{McCormick2009}
found a similar issue in an internet survey.

Our estimates for women who were raped in the past year reveal a common
issue with ARD questions. Though the questions asks respondents to
recall only women who were raped, we hypothesize respondents will
include men who are connected to a woman who was raped, even if the
woman does not meet the definition of a tie. Respondents may also be
likely to over-recall such traumatic events. Similarly, our estimates
for female suicides are consistently higher than for males (though the
difference is well within the uncertainty of measurements). Males are
actually nearly four times as likely to commit suicide as
females [\citet{UntitledQao55G31}]. This discrepancy might be
because of the isolation of many suicide victims before their deaths,
making them difficult to reach with ARD. Our estimates are especially
consistent with the case of males being more isolated than females
before committing suicide.

\begin{figure}

\includegraphics{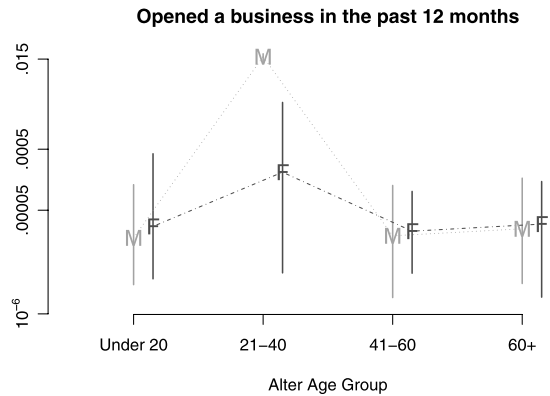}

\caption{Estimates of latent profiles for individuals starting their
own business. Letters correspond to posterior medians, while lines
represent the width of the middle half of the posterior distribution.
The lighter text represents males and the darker text females. Overall
estimates are higher for males than for females, with the largest
discrepancy for young adults.}
\label{figsmallb}
\end{figure}

Recent work has used ARD for estimating population-level social
phenomenon outside the context of hard-to-reach groups [\citet
{Detal09}]. To demonstrate the applicability of latent profile
estimation in this context, Figure~\ref{figsmallb} shows the latent
profile of individuals who opened a small business in the past year.
The trend across ages in the profiles for males and females is similar,
with most new business openers being younger adults [\citet
{smallb}]. The fraction of males opening a business is consistently
higher, however. This discrepancy is especially pronounced among young
adults, the group with highest overall propensity.
%

%

Overall, our estimates of latent profiles are similar to estimates from
other sources for the U.S. population.
The similarity between previous knowledge about the profiles of these
populations and our estimates indicates that ARD contain a significant
amount of information about the latent structure of these subpopulations.
The estimates presented in this section were obtained using the MCMC
algorithm described in Section~\ref{secmodel}. In the following section
we present an alternative regression-based estimation strategy which is
significantly less time-consuming to implement and provides comparable
performance when certain conditions are satisfied.
%
\section{Simple calculations and design recommendations}
\label{secsimple}
Given data from an existing survey, we have shown that our method will
recover features of unobserved subpopulation profiles. We propose an
alternative strategy to recover this information under certain
conditions without using MCMC.
Our simple method combines estimation and survey-design strategy,
making it well-suited for researchers who intend to collect ARD.
\citet{MSZ08} proposed the \textit{scaled-down} condition for
selecting subpopulations to reduce bias in simple estimates of
respondent degree. To estimate latent profiles, we need accurate degree
and mixing matrix estimates. To accurately estimate the mixing matrix,
we introduce a missing data perspective for ARD and propose an
estimator based on the EM algorithm.

%
In Section~\ref{secdeg} we review degree estimation and the scaled-down
condition.
Next, Section~\ref{secssc} describes a simple ratio estimator for the
mixing matrix motivated by the EM algorithm. We then describe a
regression based estimator for latent profiles in Section \ref
{secreghp} and demonstrate its effectiveness through simulation studies
in Section~\ref{secsim}.
%
\subsection{Estimating degree}
\label{secdeg}
In this section we review work on estimating respondent degree
by \citet{MSZ08}. We use these estimates in subsequent sections to
estimate mixing rates and latent profiles.

\citet{MSZ08} develop a degree estimator based on the scale-up
method of \citet{PKetal98soc}. This approach uses respondents'
answers to ARD questions and recalibrates based on the proportion of
the total population comprised of the populations used on the survey.
For example, if a respondent reports knowing 3 women who gave birth,
this represents about 1-millionth of all women who gave birth within
the last year. This information then could be used to estimate that the
respondent knows about 1-millionth of all Americans,
$(3/3.6\mbox
{ million} )\cdot(300\mbox{ million}) \approx250\mbox{ people}$.

The precision of this estimate can be increased by averaging responses
of many groups, yielding the scale-up estimator [\citet{PKetal98soc}],
\[
\hat{d}_{i}=\frac{\sum_{k=1}^{K}y_{ik}}{\sum_{k=1}^{K}N_{k}}\cdot N,
\]
where $y_{ik}$ is the number of people that person i knows in
subpopulation $k$, $N_{k}$ is the size of subpopulation $k$, and $N$ is
the size of the population.

The scale-up estimator is easy to compute, yet can induce substantial
bias if subpopulations aren't selected correctly. The scale-up
estimator assumes \textit{random mixing} across the $K$ populations. That
is, that the propensity for an individual to know members of a
subpopulation depends only on the size of the subpopulation. In
practice, this is rarely the case, as individuals tend to know more
alters who are demographically similar to themselves.

\citet{MSZ08} derived a \textit{scaled-down} condition for selecting
names so that the collection of individuals with first names that are
used to collect ARD constitute a balanced and representative sample of
the population. In other words, the combined demographic profiles of
the used first names match those of the general population.
Specifically,
\[
\frac{\sum_{k=1}^K N_{ak}}{N_{a}} = \frac{\sum_{k=1}^K N_k}{N}\qquad \forall a.
\]
Using the scaled-down condition, \citet{MSZ08} demonstrate that
the scale-up estimator produces reduced-bias estimates of degree. In
deriving the subsequent latent profile estimates, we assume we have
selected subpopulations which satisfy the scaled-down condition.
%
\subsection{A simple ratio estimator of individual mixing rates}
\label{secssc}
If for a given respondent, $i$, we could take all the members of the
social network with which $i$ has a link and place them in a room, we
would compute the mixing rate between the ego and a given alter group,
$a=(1,\ldots,A)$, by dividing the room in $A$ mutually exclusive sections
and asking alters to stand in their respective group. The estimated
mixing rate would then be the number of people standing in a given
group divided by the number of people in the room.

We could also perform a similar calculation by placing a simple random
sample of size $n$ from a population of size $N$ in a room. Then, after
dividing the alters into mutually exclusive groups, we could count
$y_{ia}$ or the number of alters respondent $i$\vadjust{\goodbreak} knows in the sample who
are in each of the $a$ alter groups. Since we have a simple random
sample, we can extrapolate back to the population and estimate the
degree of the respondent, $\hat{d}_{i}$, and within alter group degree,
$\hat{d}_{ia}$, as
%
\[
\hat{d}_{i}=\sum_{a=1}^{A}y_{ia}/({n/N}) \quad\mbox{and}\quad
\hat{d}_{ia}={y_{ia}}/({n_{a}/N_{a}}).
\]
Given these two quantities, we can estimate the mixing rate between the
respondent and an alter group by taking the ratio of alters known in
the sample who are in alter group $a$ over the total number known in
the sample. This computation is valid because we assumed a simple
random sample, thus that (in expectation) the demographic distribution
of alters in our sample matches that of the population.

In ARD, the distribution of the hypothetical alters we sample depends
on the subpopulations we select. If we only ask respondents
subpopulations which consist of young males, for example, then our
hypothetical room from the previous example would contain only the
respondent's young, male alters. Estimating the rate of mixing between
the respondent and older females would not be possible in this
situation. Viewed in this light, ARD is a form of cluster sampling
where the subpopulations are the clusters and respondents report the
presence/absence of a tie between all alters in the cluster. Since the
clusters are no longer representative of the population, our estimates
need to be adjusted for the demographic profiles of the
clusters [\citet{L99}]. Specifically, if we observe $y_{ika}$ for
subpopulations $k=(1,\ldots,K)$ and alter groups $a=(1,\ldots,A)$, then our
estimates of $\hat{d}_{i}$ and $\hat{d}_{ia}$ become
\[
\hat{d}_{i}={\sum
_{k=1}^{K}y_{ik}}\bigg/ \Biggl({\sum
_{k=1}^{K}N_{k}/N} \Biggr)
\quad\mbox{and}\quad
\hat{d}_{ia}={\sum_{k=1}^{K}y_{ika}}\bigg/
\Biggl({\sum_{k=1}^{K}N_{ak}/Na}
\Biggr),
\]
where $N_{k}$ is the size of subpopulation $k$ and $N_{ak}$ is the
number of members of subpopulation $k$ in alter group $a$. To estimate
the mixing rate, we could again divide the estimated number known in
alter group $a$ by the total estimated number known. Under the
\textit{scaled-down} condition the denominators in the above expressions
cancel and the mixing estimate is the number known in the
subpopulations that are in alter group $a$ over the total number known
in all $K$ subpopulations.

In the examples above, we have assumed the alters are observed so that
$y_{ika}$ can be computed easily. This is not the case in ARD, however,
since we observe only the aggregate number of ties and not the specific
demographic makeup of the recipients. Thus, ARD are a cluster sample
where the specific ties between the respondent and members of the alter
group are \textit{missing}.

If we ignore the residual variation in propensity to form ties with
group $k$ individuals due to noise [see (\ref{eqrecallmodel})
in\vadjust{\goodbreak}
Section~\ref{secmodel}], we may assume that the number of members of
subpopulation $k$ in alter group $a$ the respondent knows, $y_{ika}$,
follows a Poisson distribution. Under this assumption, we can estimate
$m_{ia}$ by imputing $y_{ika}$ as part of an EM algorithm [\citet
{EM}]. Specifically, for each individual define $\mathbf
{y}_{ik}^{(\mathrm{com})}=(y_{ika},\ldots, y_{i1A})^{T}$ as the
complete data vector for each alter group. The complete data
log-likelihood for individual $i$'s vector of mixing rates, $\mathbf
{m}_{i}=(m_{i1},\ldots,m_{iA})^{T}$, is $\ell(\mathbf{m}_{i};
\mathbf{y}_{i1}^{(\mathrm{com})},\ldots,\mathbf{y}_{iK}^{(\mathrm{com})})$,
which has the form
%
\begin{eqnarray}\label{eqlik}
&&\ell\bigl(\mathbf{m}_{i}; \mathbf{y}_{i1}^{(\mathrm{com})},\ldots,
\mathbf {y}_{iK}^{(\mathrm{com})}\bigr)\nonumber\\[-8pt]\\[-8pt]
&&\qquad=\sum
_{k=1}^{K}\sum_{a=1}^{A}
\log \biggl(\mbox {Poisson} \biggl(y_{ika}; \lambda_{ika}=d_{i}m_{ia}
\frac
{N_{ak}}{N_{a}} \biggr) \biggr).\nonumber
\end{eqnarray}
%
%
%
Using (\ref{eqlik}), we derive the following two updating steps for
the EM:
\begin{eqnarray*}
{y}_{iak}^{(t)}&=&y_{ik} \biggl(\frac{m_{ia}^{(t-1)}
({N_{ak}}/{N_{a}})}{\sum_{a=1}^{A}m_{ia}^{(t-1)}
({N_{ak}}/{N_{a}})}
\biggr),
\\
{m}_{ia}^{(t)}&=&\frac{\sum_{k=1}^{K}y_{ika}^{(t-1)}}{\sum_{k=1}^{K}y_{ik}}.
\end{eqnarray*}
If one sets $m_{ia}^{(0)}=N_{a}/N$, which corresponds to random mixing
in the population, and runs one EM update, this would result in the
following \textit{simple ratio estimator} of the mixing rate for
individual $i$:
%
\begin{equation}\label{eqsimple}
\hat{m}_{ia}=\frac{\sum_{k-1}^{K}y_{ik}(N_{ak}/N_{k})}{\sum_{k-1}^{K}y_{ik}}.
\end{equation}
%
In our simulation studies, this simple estimator produces estimates
very close to the converged EM estimates. Additionally, it is easy to
show that the simple ratio estimate, $\hat{m}_{ia}$, is unbiased if
$N_{ak}/N_{a} \neq0$ for only one alter group $a$ and that for any $a$
there exists a subpopulation, $k$, such that $N_{ak}=N_{a}$. We refer
to this condition as \textit{complete separability}. Therefore, (\ref
{eqsimple}) constitutes a simple estimate for individual mixing rate
and can be used to estimate average mixing behaviors of any ego group.

\subsection{Regression-based estimates for latent profiles}
\label{secreghp}

The estimates for respondent degree and mixing estimates rely on latent
profile information from some populations. Using these estimates, we
now develop a regression-based estimator for unobserved latent profiles.
For each respondent and each unknown subpopulation we now have
%
\begin{equation}\label{eqregression}
y_{ik}=\sum_{a=1}^A
\hat{d}_i\hat{m}_{ia}h(a,k).
\end{equation}
If we\vspace*{1pt} denote $\mathbf{X}_k$ as the $n \times A$ matrix with elements
$\hat{d}_{i}\hat{m}_{ia}$ and the vector $h(\cdot,k)=\overrightarrow
{\beta_k}$, then (\ref{eqregression}) can be regarded as a linear
regression equation, $\overrightarrow{y}_k=\mathbf
{X}_k\overrightarrow
{\beta_k}$, with the constraint that coefficients, $\overrightarrow
{\beta_k}$, are restricted to be nonnegative. 
\citet{ls} propose an algorithm for computing these coefficients.
Since the $\hat{m}_{\cdot k}$ sum to one across alter groups, the
columns of $\mathbf{X}_k$ are collinear. This could produce instability
in solving the quadratic programming problem associated with finding
our estimated latent profiles. In practice, we have found our estimates
perform well despite this feature.

\subsection{Simulation experiments}
\label{secsim}

We present simulation experiments to evaluate our regression-based
estimates under four strategies for selecting observed profiles. First,
we created profiles which are completely separable. Second, we
constructed profiles for the names satisfying the scaled-down condition
presented in \citet{MSZ08} using data from the Social Security
Administration. These names provide insights into the potential
accuracy of our method using actual profiles. As a third case, we
include the names from \citet{MSZ08} which violate the
scaled-downed condition and are almost exclusively popular among older
respondents. For the fourth set of names, recall from Section
\ref{secmodel} that the mixing matrix estimates are identifiable only if
the matrix of known profiles, $\mathbf{H_{A\times K}}$, has rank $A$.
To demonstrate a violation of this condition, we selected a set of
names with uniform popularity across the demographic groups, or nearly
perfect collinearity. There is some correlation in the scaled-down
names since several names have similar profiles. The degree of
correlation is substantially less than in the flat profiles, however.

In each simulation, we generated 500 respondents using the Latent
Nonrandom Mixing Model in (\ref{eqrecallmodel}) [see also \citet
{MSZ08}] with each of the four profile strategies. Mixing matrix
estimates were calculated using the simple estimate derived from the
first step of the EM algorithm in Section~\ref{secssc}. We compare our
mixing matrix estimates to the estimated mixing matrix from
\citet{MSZ08}, which we use to generate the simulated data. We evaluate the
latent profiles using six names with profiles known from the Social
Security Administration. We repeated the entire process 1000 times.
%
\begin{figure}

\includegraphics{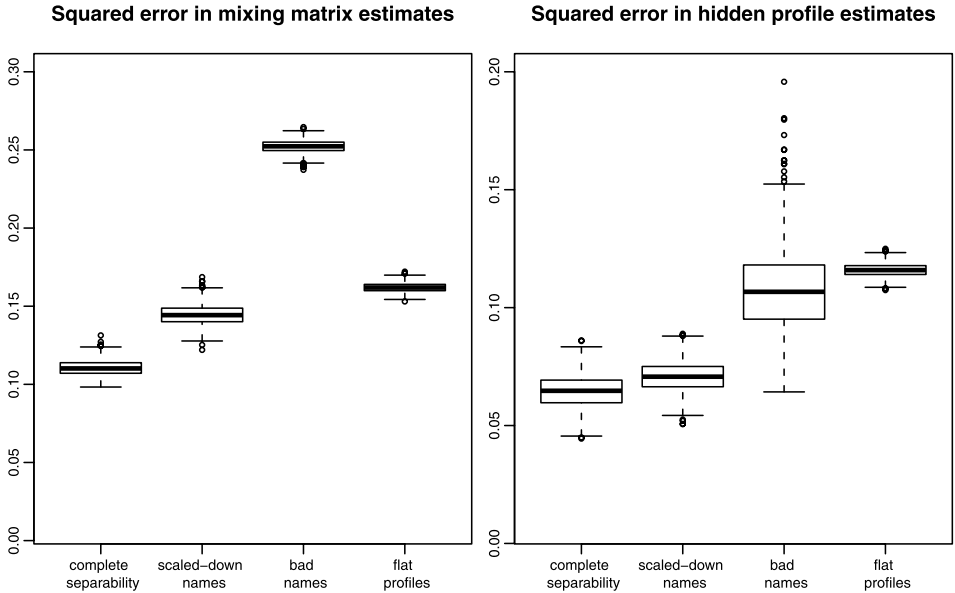}

\caption{Total mean squared error across all elements of the mixing
matrix and latent profile matrix. The vertical axis is the sum of the
errors across all eight alter groups. We generated 500 respondents
using the four profile structures, then evaluated our ability to
recover the mixing matrix estimated in McCormick, Salganik and Zheng
(\protect\citeyear{MSZ08}) and the known
profiles of six additional names. We repeated the simulation 1000
times. In both cases the ideal profile has the lowest error, followed
by the scaled-down names suggested by McCormick, Salganik and Zheng
(\protect\citeyear{MSZ08}).}
\label{fighidden}\vspace*{-3pt}
\end{figure}
Figure~\ref{fighidden} presents boxplots of the squared error in mixing
matrix and latent profile estimates. In both cases, the ideal,
completely separable, profiles have the lowest error. The scaled-down
names also perform well, indicating that reasonable estimates are
possible even when complete separability is not. The flat profiles
perform only slightly worse than the scaled-down names for estimating
mixing but significantly worse when estimating latent profiles. The
names which violate the scaled-down condition produce poor estimates of
both quantities.
%

\section{Conclusion}
We present a method for estimating latent profiles in hard-to-reach
groups using standard surveys. Our method has two stages. First, we use
known profiles for some populations to estimate respondent degree and
the rate of mixing between survey respondents and groups in the
population. Next, conditional on these estimates, we infer latent
structure in populations where profiles are unknown. For existing data,
we present a Bayesian hierarchical model and MCMC algorithm. We also
propose viewing ARD in the context of missing data and provide a simple
ratio estimate of mixing rates based on the EM algorithm. We then
describe a regression-based estimate for latent profiles.

Despite its utility, there are several known issues with ARD. Using ARD
in hard-to-reach populations presents special challenges which
intersect with these known issues. Many events in this context are
especially traumatic, leaving a more persistent signal in the
respondent's memory than a typical tie. This phenomenon causes
respondents to over-count their ties with a specific subpopulation. In
Section~\ref{secdata} we contend that our overestimation of the
proportion of men who are women who were raped in the past year is due
to respondents overestimating by counting males who are associated with
females who have been raped, for example. This issue is in some sense
the opposite of that faced by early ARD surveys for degree estimation
when the concern was respondents under-recalling acquaintances from
large populations [\citet{PKetal03}]. Hard-to-reach groups are
also often more open to interpretation than standard subpopulations.
\citet{CMetal01} give the example of people opening their own
business and the homeless, for example. While there is some ambiguity
in whether or not an individual has opened a new business, there is
likely much greater variability between respondents in their
classification of an individual as homeless. Hard-to-reach groups are
also often associated with social stigma. This stigma increases the
likelihood that a respondent will know a member of a subpopulation but
not be aware that the alter belongs to the subpopulation, known as
transmission errors. Recent work by \citet{contacts} offers new
insights into the magnitude of transmission errors in the context of
HIV/AIDS, though the nature of the error likely depends heavily on the
specific group of interests (respondents' decisions to reveal HIV
status are likely quite different than their decision to discuss
diabetes, e.g.).

This method also makes an assumption that the networks of the
respondents are representative of networks of similar individuals in
the population. In Section~\ref{secdata}, in our discussion of the
ratio of males to females who commit suicide, another possible
explanation is that our survey does not include enough individuals who
are likely to know people who commit suicide. This bias could be
present in the networks of respondents even if the sample is, within
respondents, observably representative. This point demonstrates the
potential for future work in modeling bias that comes not from the
respondents selected, but from the features of the networks of these
respondents. This type of sampling bias is related to previous work
by \citet{Lavallee2007uz} and could prove a promising area for
future work.

Our method demonstrates that ARD capture aspects of latent social
structure through indirect observations of the social network. To do
this, however, we require known profiles for some subpopulations. This
requirement limits the estimable latent profiles to features which are
known for some subpopulation. In our examples we use first names and
estimate age and gender profiles. We may, for example, be interested in
the race/ethnic profiles of the hard-to-reach populations. We are
unable to estimate this from our current data because of the issues
with obtaining demographic profiles for first names mentioned in
Section~\ref{secdata}. An alternative approach, and direction for
potential future work, would be estimating a geometric,
multidimensional latent social space based on features of the actors
and the social network [\citet{HRH02,H05}]. Such a technique would
provide a sense of the broad topography of the network (similar to
Bayesian multi-dimensional scaling) and elucidate similarities between
network structure in hard-to-reach groups.

An additional direction for future work involves combining information
from ARD with other forms of data collection to better understand
hard-to-reach groups. As mentioned in Section~\ref{secintro}, RDS
provides detailed information about a biased sample of members of the
hard-to-reach group. This detailed information is in contrast to the
indirect, general information obtained through ARD. The missing-data
framework presented in Section~\ref{secsimple} provides a first-step
toward a general framework for combining information across various
network-based data collection strategies.

\section*{Acknowledgments}

The authors gratefully acknowledge the support of the
National Science Foundation, the Columbia Applied Statistics Center,
and the Columbia Population Research Center, along with the helpful
comments of the Editor, Associate Editor, and an anonymous Referee.



\printaddresses

\end{document}